\documentclass[aip,jap,reprint,a4paper,superscriptaddress]{revtex4-1}

\usepackage{graphicx}
\usepackage{mdframed}
\usepackage{natbib}
\usepackage{amsmath}
\usepackage{amssymb}
\usepackage{color}
\usepackage{esint}
\usepackage{subfigure}
\usepackage{soul}
\usepackage{color}
\usepackage{xcolor}
\usepackage{bm}
\definecolor{Red}{rgb}{0.9,0.0,0.1}
\usepackage[colorlinks=true,citecolor=blue,linkcolor=blue,hyperindex]{hyperref}
\usepackage{lineno}

\begin{document}


\title{Controlling energy transfer time between two coupled magnetic vortex-state disks}

%
\author{H. Vigo-Cotrina}
\affiliation{Centro Brasileiro de Pesquisas F\'{\i}sicas, 22290-180,  Rio de Janeiro, RJ, Brazil}

\author{A.P. Guimar\~aes}
\affiliation{Centro Brasileiro de Pesquisas F\'{\i}sicas, 22290-180,  Rio de Janeiro, RJ, Brazil}

\date{\today}
\begin{abstract}
The influence of the in-plane uniaxial anisotropy (IPUA) in the mutual energy transfer time ($\tau$) between two identical coupled nanodisks was studied. Using an analytical dipolar model we obtained the  interactions between the disks along x and y directions (the coupling integrals) as a function of the uniaxial anisotropy constant (K$_{\sigma}$) and the distance. We find that the IPUA increases the interaction between the disks allowing shorter energy transfer times. For our range of K$_{\sigma}$ values we get a drop in the values of $\tau$ of up to about 70$\%$. From the lagrangian  of the system we obtained the equations of motion and the coupling frequencies of the dynamic system as a function of distance and K$_{\sigma}$. The coupling frequencies were also obtained from micromagnetic simulations. Our results of the simulations are in agreement with the analytical results.

\end{abstract}
\maketitle

\section{Introduction}

Depending on their geometries and sizes, magnetic nanostructures can present a vortex configuration as their ground states.\cite{Guslienko:2008, Guimaraes:2009} This configuration is characterized by curling magnetization in the plane of the nanostructure, and a core where the magnetization points out of the plane. The curling direction defines the circulation $C = +1$ (counterclockwise (CCW)) and $C = -1$ (clockwise (CW)). The core has polarity $p_0 = +1$ when it points in the $+z$ direction and $p_0 = -1$ in the $-z$ direction. The magnetic vortices have several potential applications for devices, such as media for magnetic random access memories (MRAM), spin torque induced magnetization processes, microdisks for targeted cancer-cell destruction, etc.\cite{Guimaraes:2009,Locatelli:2014, Bohlens:2008, Belanovsky:2013}\\
\indent When vortices are excited from their equilibrium position and allowed to relax, they perform a motion called gyrotropic, with an eigenfrequency (gyrotropic frequency)  in the sub-gigahertz range.\cite{Guimaraes:2009, Guslienko:2008, Novosad:2005, Guslienko:2006, Novosad:2002} This eigenfrequency depends on the ratio of the thickness to the radius of the disk ($\beta = L/R$).\cite{Guimaraes:2009, Guslienko:2006} \indent The control of gyrotropic frequency values has been reported in previous works using perpendicular magnetic fields and polarized spin current.\cite{Yoo:2011, Choi:2008, Loubens:2009} Nevertheless, problems such as the Joule heating and stray magnetic fields are undesirable for device applications\cite{Cavill:2013}, therefore a new mechanism for controlling the frequencies is needed. In this sense the influence of perpendicular uniaxial anisotropy (PUA) and in-plane uniaxial anisotropy (IPUA) in the magnetization process of magnetic vortices has gained interest in recent years.\cite{Roy:2013, Ostler:2015, Parreiras:2015, Novais:2011, Novais:2013, Garcia:2010, Fior:2016} In multilayer systems a perpendicular uniaxial anisotropy (PUA) can be induced varying the substrate thickness\cite{Garcia:2010}, while IPUA can be induced through inverse magnetostriction effect by voltage-induced strain via a PZT piezoelectric transducer\cite{Parkes:2013, Ostler:2015}. The IPUA allows real-time control of the eigefrequency in an appreciable range\cite{Roy:2013}. The eigenfrequency can also be controlled using PUA, as has already been demonstrated by Fior \textit{et al.} \cite{Fior:2016}, but the downside is that this control cannot be made in real-time, and the range of frequency variation is barely 3$\%$ before the vortex is destabilized.\\
\indent When magnetic disks are close to one another, there arises a frequency splitting due to the magnetic interaction\cite{Shibata:2003}. Expressions for the magnetic vortex excitation frequencies and coupling integrals in a pair of coupled circular disks of equal radii were obtained by Shibata \textit{et al.}\cite{Shibata:2003} and Sukhostavets \textit{et al.}\cite{Sukhostavets:2011} These expressions were also obtained for the case of coupled circular disks of different radii by Sinnecker \textit{et al.}\cite{Sinnecker:2014}\\
\indent Coupling different disks with magnetic vortices allows the possibility of loss-less energy transfer between them\cite{Jung:2011}, and the propagation of the information;\cite{kim:2012} which is relevant for the flow of information in devices using magnetic vortices. 
The energy transfer time of a disk to the other is caracterized by the parameter $\tau$. This parameter is inversely proportional to the splitting frequency ($\bigtriangleup\omega/2\pi$).\cite{Jung:2011} In this sense, the control of the mutual energy transfer time $\tau$ is very important to characterize this transfer.

\indent The goal of this work is to propose a novel method for controlling $\tau$ in a pair of identical nanodisks, using the influence of the IPUA.  To date there is only one method of controlling the value of $\tau$, using perpendicular magnetic fields\cite{Han:2012}. We used micromagnetic simulations  and found a correlation between the IPUA  values and $\tau$. The coupling integrals Ix and Iy between disks  and the splitting frequency  are also affected by the presence of the IPUA. In order to gain physical meaning and obtain the new values of the coupling integral and splitting frequency, a simple analytic method considering dipolar interaction was used. Our analytical results are in accordance with the micromagnetic simulations.\\
\indent The micromagnetic simulations were made using the open source software Mumax\cite{Vansteenkiste:2014}. The magnetoelastic energy was included in the micromagnetic simulation as a uniaxial anisotropy energy, as proposed in reference 13. We used cells of  2 $\times$ 2 $\times$ 7\:nm$^3$. The magnetostrictive material used was Galfenol (FeGa), which is of great interest and exhibits high magnetostriction\cite{Parkes:2013}. The material parameters of Galfenol used here were\cite{SungHwan:2010, Summers:2007, Roy:2013}: saturation magnetization $M_s = 1.360 \times 10^6$\:A/m$^2$, exchange stiffness $A = 14 \times 10^{-12}$ J$/$m. We used gyromagnetic ratio $\gamma =$ 2.21 $\times$ $10^5$\:m$/$As, damping constant $\alpha = 0.01$ and IPUA ranging from 0 to 58500 J$/$m$^3$. For larger values of the anisotropy, the magnetic vortex configuration is not stable. We used two identical disks, located along the x-axis, with diameters 256\:nm, thickness L $=$ 7\:nm and  separation distance D between the disks (Fig. \ref{fig:discos}). All our simulations were made considering uniaxial anisotropy in the direction of the x axis .

\section{Results and discussion}
\label{section:results}

\subsection{Micromagnetic simulations}
\label{subsection:micromagneticsimulations}
The gyrotropic frequency variation depending on the induced uniaxial anisotropy constant K$_{\sigma}$ for an isolated disk has already been studied in detail by Roy\cite{Roy:2013} for the same geometry that we used, however, we will make a brief description of this case. 

\begin{figure}[h]
\centering
\includegraphics[width=1\columnwidth]{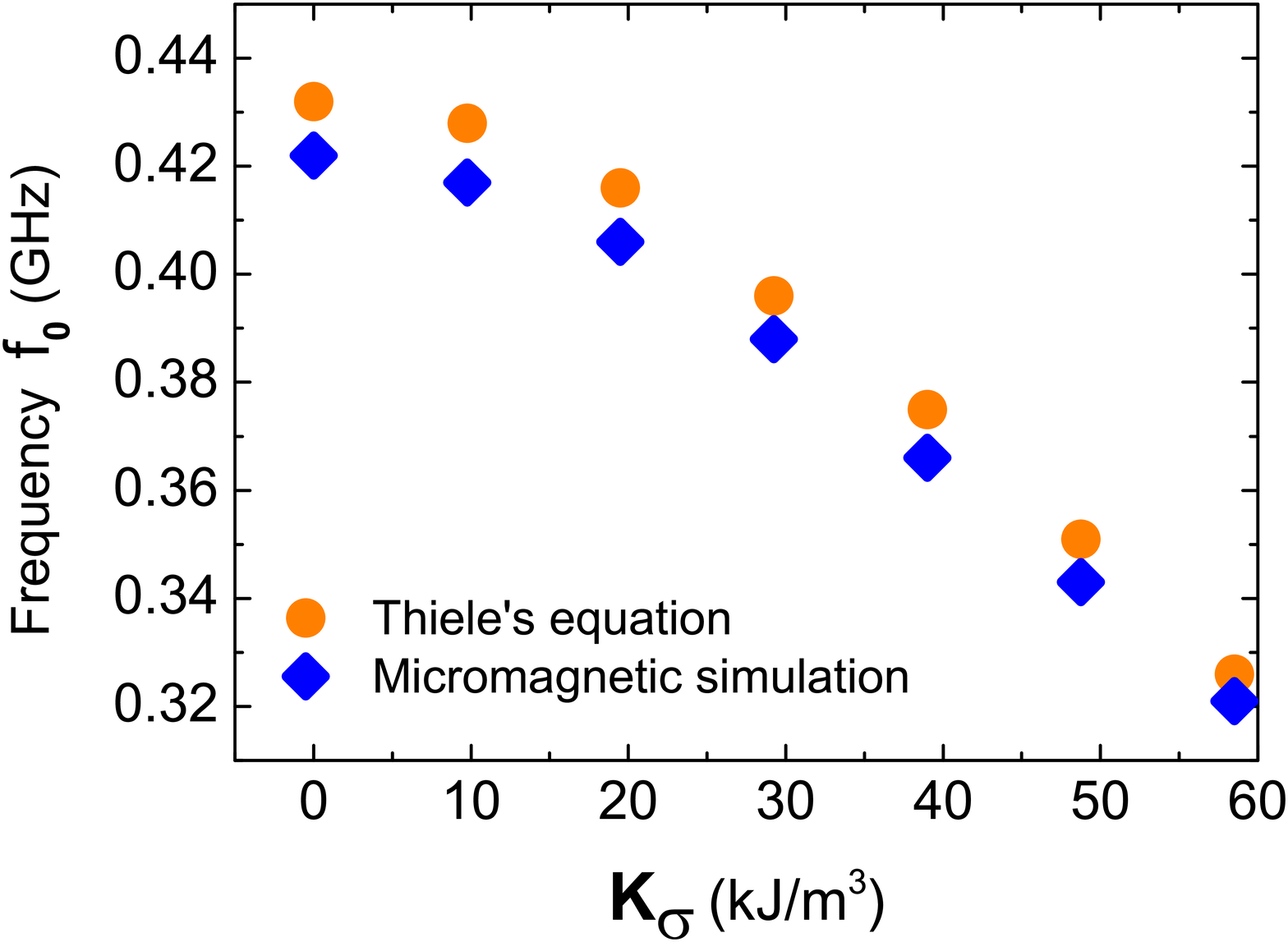}
\caption{Gyrotropic frequency variation with respect to K$_{\sigma}$ for a isolated disk of diameter 2R = 256\: nm and thickness L = 7\: nm. 
Blue diamonds and orange dots represent the values obtained from micromagnetic simulations, and  Thiele's equation ($\kappa(2\pi G)^{-1}$), respectively.}\label{fig:eigenfrequency}
\end{figure}

The vortex core is initially at the equilibrium position, at the center of the disk. In order to induce gyrotropic movement, we first apply a static field of 20\: mT in the +x direction for a few nanoseconds   using a large damping $\alpha$ = 1 for faster convergence, then this field is turned off and a typical damping $\alpha$ = 0.01 was used, allowing the vortex core to perform the gyrotropic motion. The eigenfrequency (f$_0$ = $\omega_0/2\pi$) is obtained by a fast Fourier transform (FFT) from the time evolution of the magnetization. The frequency decrease with increasing anisotropy, as shown in Fig. \ref{fig:eigenfrequency}.

 \begin{figure}[h]
\centering
\includegraphics[width=1\columnwidth]{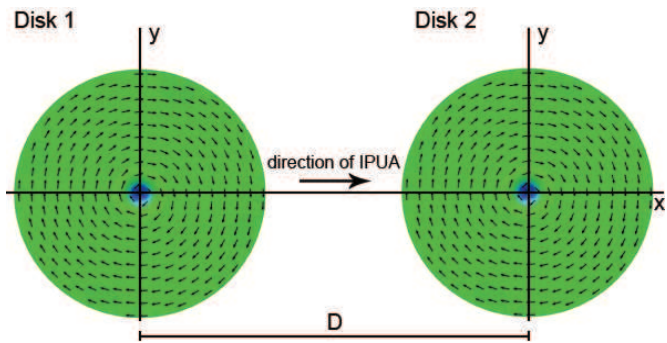}
\caption{Coupled disks with magnetic vortex configuration separated by a center-to-center distance D.}\label{fig:discos}
\end{figure}

The gyrotropic frequencies can also  be determined analytically using the linearized Thiele's equation\cite{Thiele:1973}. Considering a small damping, this equation can be written as:

\begin{eqnarray}
\textbf{G} \times \frac{d\textbf{X}}{dt} - \frac{\partial W (\textbf{X})}{\partial \textbf{X}} = 0,\label{eq:Thiele}
\end{eqnarray}

\noindent where $\textbf{G}$ is the gyrovector $\textbf{G} = - Gp_0\hat{z}$, $G = 2\pi \mu_0M_sL/\gamma$ is the gyrotropic constant (it is assumed that the magnetization does not vary along the thickness of the disk, an assumption valid for $\beta$ = L/R $<<$ 1), $\gamma$ is the gyromagnetic ratio and $M_s$ is the saturation magnetization; $W (\textbf{X}) = W(0) + \frac{1}{2}\kappa\textbf{X}^2 $ is the potential energy and $\kappa = 40\pi M_s^2L^2/9R$ is the stiffness coefficient calculated within the side-charge-free model at $\beta$ $<<$ 1 \cite{Guslienko:2008}, and R is the disk radius. The gyrotropic frequency is given by the expression $f_0 = \kappa(2\pi G)^{-1}$. The stiffness coefficient and the gyrotropic constant were obtained from micromagnetic simulations following the methodology of some previous works.\cite{Han:2012, Roy:2013, Yu:2006} Thus, the gyrotropic constant was obtained using the expression\cite{Yu:2006}:

\begin{equation}
G = \frac{L}{\gamma M_s^2}\int_S \bm{M}.[(\frac{\partial \bm{M}}{\partial x}) \times (\frac{\partial \bm{M}}{\partial y})] dS
\end{equation}

\begin{figure}[h]
\centering
\includegraphics[width=1\columnwidth]{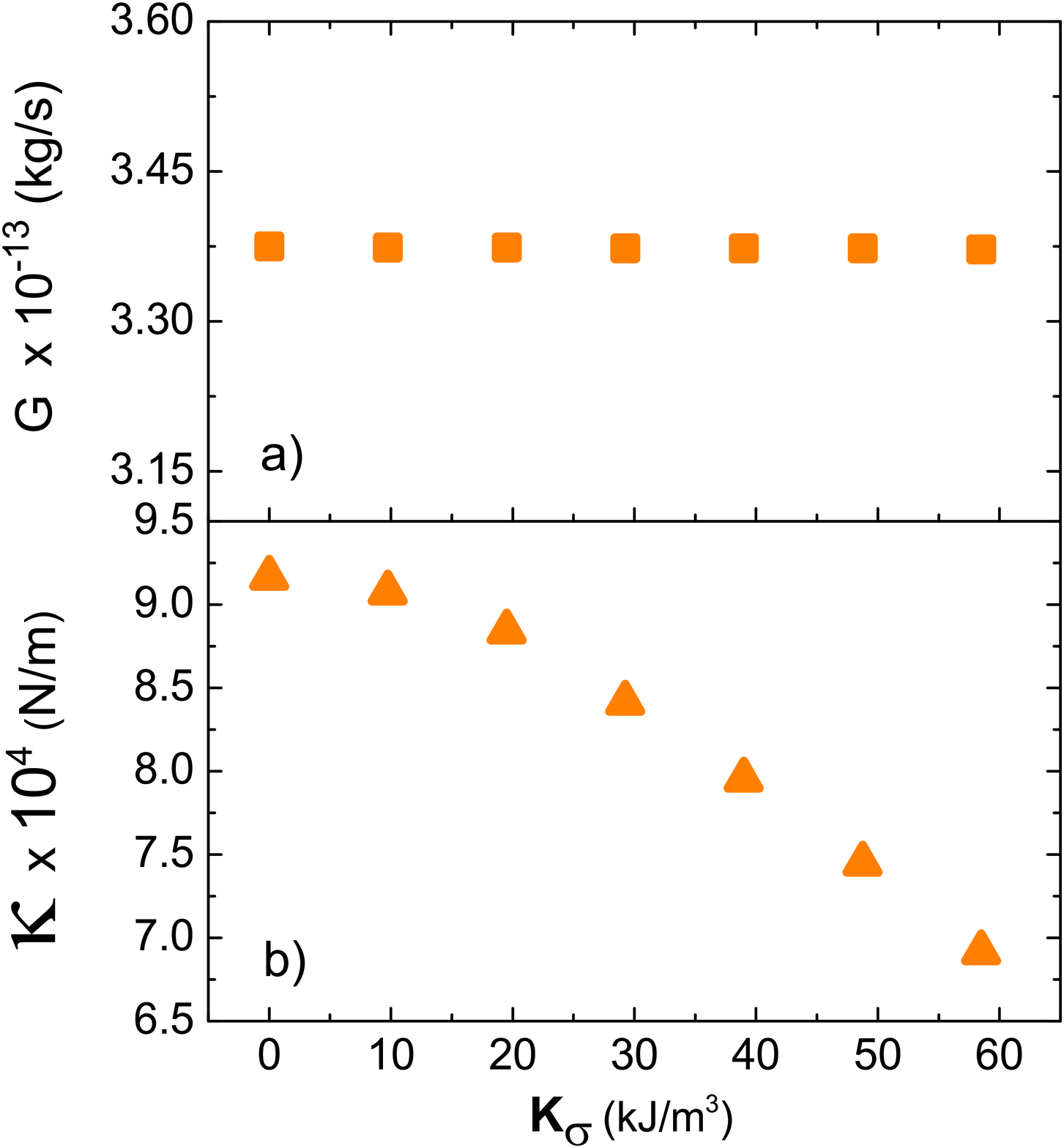}
\caption{a) Gyrotropic constant G and b) stiffness coefficient $\kappa$ obtained from micromagnetic simulations.}\label{fig:constantesthiele}
\end{figure}

\noindent and the stiffness coefficient was obtained from the slope of the linear fits of $W(\bm{X})$ versus $\bm{X}^2$.\cite{Roy:2013, Yoo:2011}
The values of the gyrotropic constant and stiffness coefficient obtained from the analytical expressions are G = 3.401$\times 10^{-13}$\:kg/s  and $\kappa$ = 9.86\:N/m for K$_{\sigma}$ = 0\:kJ/$m^3$ while the values obtained from micromagnetic simulations are 3.375$\times 10^{-13}$\:kg/s and $\kappa$ = 9.16\:N/m. There is a good agreement between the results obtained from the analytical expressions and those obtained from micromagnetic simulation. With the presence of the IPUA the value of the gyrotropic constant remains unchanged, as shown in Fig. \ref{fig:constantesthiele}, which means that the IPUA does not alter the profile of the vortex core\cite{Roy:2013}, whereas the  stiffness coefficient  shows a falling value with increasing anisotropy constant K$_{\sigma}$. This fall is due to the competition between exchange and demagnetizing energy versus magnetoelastic energy\cite{Roy:2013}.
Decreasing kappa values are reflected in the values of the eigenfrequencies, as shown in Fig. \ref{fig:eigenfrequency}. Our results obtained for isolated disks are consistent with those obtained by Roy\cite{Roy:2013}.

In order to study the dependence of $\tau$ with the IPUA, we considered a system of two coupled disks, located along the x-axis, separated by a center to center distance D, as shown in Fig. \ref{fig:discos}. Initially, both vortex cores are in the equilibrium position at the center of their respective disks. In order to induce gyrotropic movement we have applied a static field in the +x direction for a few nanoseconds only on disk 1, then this field was turned off, allowing the vortex core to perform the gyrotropic motion with decreasing amplitude, while the vortex core of disk 2 begins to perform the gyrotropic moviment with increasing amplitude, due to transfer of energy from disk 1.

\begin{figure}[h]
\centering
\subfigure{\includegraphics[width=1\columnwidth]{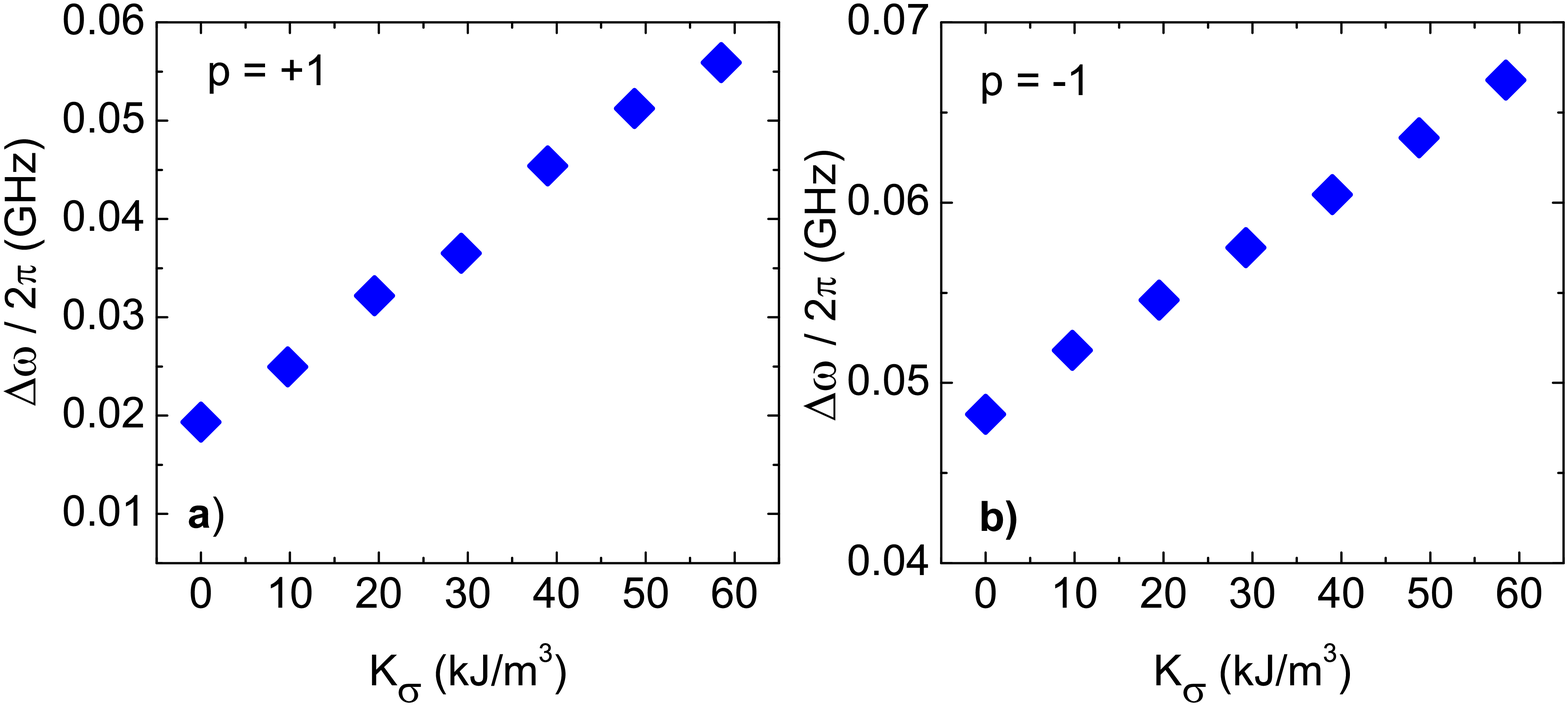}}
\subfigure{\includegraphics[width=1\columnwidth]{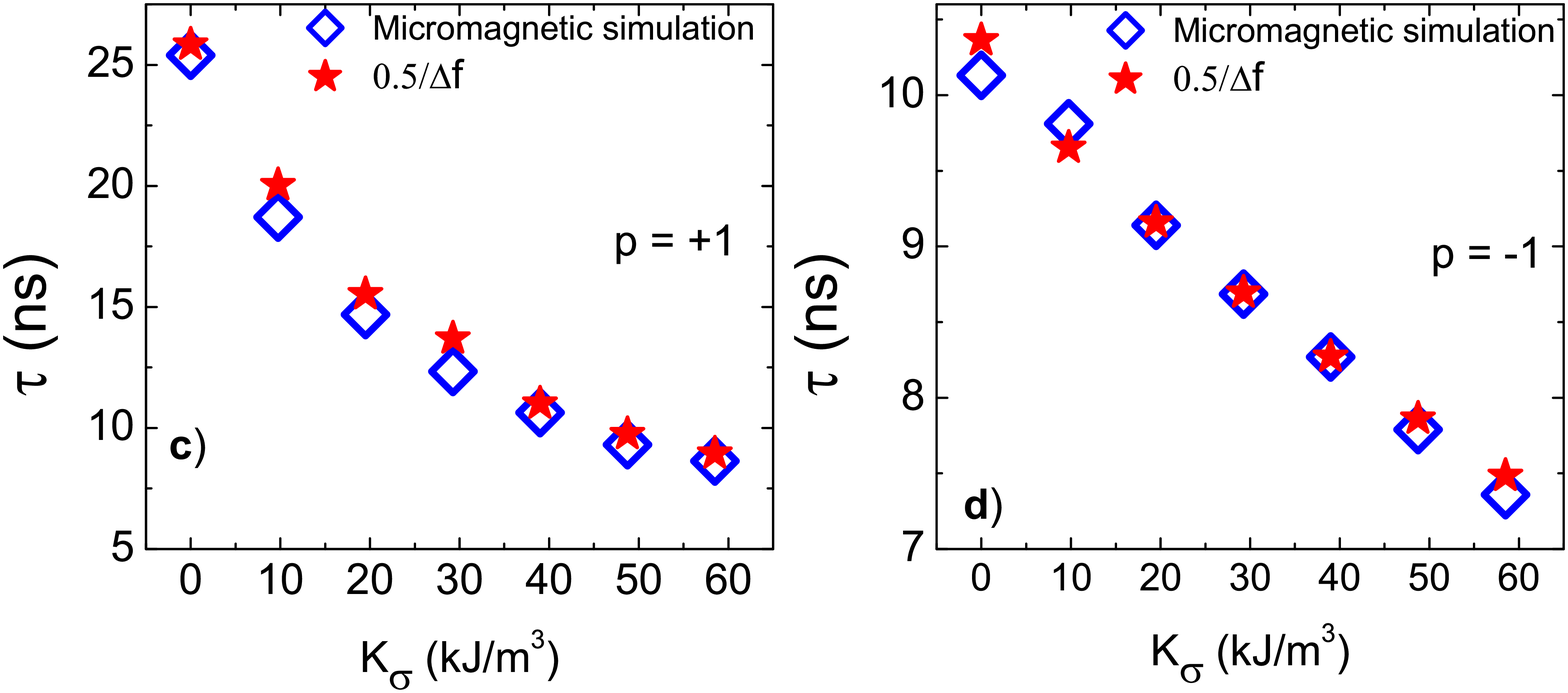}}
\caption{Splitting frequency and $\tau$ variation with respect to K$_{\sigma}$ for a) p = +1, and b) p = -1 with reduced distance d = D/R = 2.27, both obtained from micromagnetic simulations.}\label{fig:splitting}
\end{figure}

The  splitting frequency is affected by the presence of the IPUA, increasing with the increase of K$_{\sigma}$ from  19.35\:MHz (K$_{\sigma}$ = 0\:kJ/m$^3$) to 55.9\:MHz (K$_{\sigma}$ = 58.5\:kJ/m$^3$) for the case p = p$_1$p$_2$ = +1. For p = -1 the  splitting frequency increases from  48.25\:MHz (K$_{\sigma}$ = 0\:kJ/m$^3$) to  66.8\:MHz (K$_{\sigma}$ = 58.5\:kJ/m$^3$). This dependence is shown in Fig. \ref{fig:splitting}. As expected, the  splitting frequency is larger for the case p = -1 than for p = +1 because the dipolar interaction is stronger in the former case\cite{kim:2012}.

Since the  splitting frequency ($\bigtriangleup \omega$/2$\pi$) is inversely related to the energy transfer time ($\tau$)\cite{Jung:2011}, it is expected that  $\tau$ also depends on K$_{\sigma}$. This dependence is shown in Fig. \ref{fig:splitting} for p = +1 and p =-1. These values were extracted from micromagnetic simulations considering the $\tau$ definition: $\tau$ is defined as the time required by the energy of disk 1 to reach its minimum value for the first time\cite{Jung:2011}. Additionally, we have calculated $\tau$ using the expression $\tau$ = 0.5/$\bigtriangleup$f  \cite{Jung:2011}, where $\bigtriangleup$f = $\bigtriangleup\omega/2\pi$. These results are also shown in Fig. \ref{fig:splitting} .
 Another way to find the $\tau$ values is to observe the time required for envelopes of \textbf{X}$_1$ or \textbf{M}$_1$(t) reach their minimum values for the first time. In all cases, the results are almost the same. For a reduced distance d = D/R = 2.27, we found a drop in the value of $\tau$ of almost 69\%, from $\tau$ = 25.4\:ns (K$_{\sigma}$ = 0) to $\tau$ = 8.6\:ns (K$_{\sigma}$ = 58.5 kJ/m$^3$) for p = +1 and  a drop of 27\% from $\tau$ = 10.13\:ns (K$_{\sigma}$ = 0), to $\tau$ = 7.36\:ns (K$_{\sigma}$ = 58.5 kJ/m$^3$) for p = -1. For larger  d, the decrease in the value of $\tau$ with respect to the increase of  K$_{\sigma}$ remains significant (approximately 60\%) for p = +1, while for the case p = -1 the drops are reduced to 20\%.
Despite getting lower values of $\tau$ with high values of K$_{\sigma}$, it is important to note that  energy transfer times remain shorter when p = -1. This is because the coupling between the two disks is stronger in comparison with p = +1\cite{Han:2012}.
Up to this point we have shown that it is possible to control the energy transfer time using the influence  of the IPUA, thus becoming a new and effective method for controlling $\tau$. Next, we will explain the reason why $\tau$ decreases with increasing K$_{\sigma}$.

\subsection{Analytical dipolar model}
\label{section:modelodipolar}

Although the magnetic interaction between magnetic vortices depends on several multipole terms\cite{Sukhostavets:2013}, we used a simple dipolar model to understand the effect of IPUA on the coupling integrals. This model has been used in other work for the study of the magnetic interaction among  three coupled disks system\cite{Asmat:2015}.
The magnetic dipolar energy ($E_{dip}$) is given by:

\begin{equation}\label{interacciondipolar1}
E_{dip} = \frac{\mu_0}{4\pi D_{ij}^3}[\bm{\mu_i}.\bm{\mu_j} - 3(\bm{\hat{D}_{ij}}.\bm{\mu_i}).(\bm{\hat{D}_{ij}}.\bm{\mu_j})],
\end{equation}

\noindent where $\bm{\hat{D}_{ij}}$ is a unit vector along the axis connecting the centers of the disks i and j.

For small displacements, the magnetic dipolar moment for a magnetic vortex is defined by:\cite{Asmat:2015}

\begin{equation}\label{momentodipolar}
\bm{\mu_{i,j}} = -\lambda C_{i,j}M_sLR (\bm{\hat{z}}\times\bm{X_{i,j}}),
\end{equation}

with: 

\begin{equation}\label{lambda}
\lambda_{i,j} = \frac{\pi R}{M_s}\frac{\|\bm{M}_{i,j}\|}{\|\bm{X}_{i,j}\|}
\end{equation}

\noindent where $\lambda$ (in our case, $\lambda_i$ = $\lambda_j$ because the two disks have the same dimensions) is a parameter that depends on the position and the magnetization of the vortex core  which can be extracted from micromagnetic simulations, $C_{i,j}$  is the circulation of the disks i and j, $\bm{X}_{i,j} = (x_{i,j},y_{i,j})$ is the vector position of the core vortex of the disks. After some algebra and considering i, j = 1, 2 we obtain the following expression for the magnetic dipolar energy:

\begin{equation}\label{interacciondipolar2}
E_{dip} = C_1C_2(\eta^*x_1x_2 - 2\eta^*y_1y_2),
\end{equation}

\noindent where $\eta^* = \mu_0\lambda^2M_s^2L^2R^2/4\pi D_{12}^3$, $D_{12} = D$ is the center-to-center distance between 2 disks. Eq. (\ref{interacciondipolar2}) is similar to the analytical expression obtained for the magnetic interaction energy ($W_{int}$) given by:\cite{Shibata:2003,Sukhostavets:2011}.

\begin{equation}\label{interacciondipolar3}
W_{int} = C_1C_2(\eta_xx_1x_2 + \eta_yy_1y_2),
\end{equation}

Comparing Eq. (\ref{interacciondipolar2}) with Eq. (\ref{interacciondipolar3}) we obtain:

\begin{equation}\label{etas}
\eta_x = \eta^* \hspace{1cm}
\eta_y = -2\eta^*
\end{equation}

Therefore, the new coupling integrals are given by:

\begin{equation}\label{integraisdeacoplamento}
I_{x,y} = \frac{8\pi}{\mu_0RM_s^2}\eta_{x,y} 
\end{equation}

The  expressions of the coupling integral  obtained in previous works are limited to the case  when the magnetic vortex configuration is not disturbed by some external agent (e.g.: perpendicular magnetic field, perpendicular uniaxial anisotropy, IPUA). In contrast, our expressions for the coupling integrals (Eq. \ref{integraisdeacoplamento}) can be used when the magnetic vortex configuration is disturbed, as they depend on the $\lambda$ parameter, that is unique for each level of disturbance. In order to obtain the $\lambda$ parameter, we follow the methodology used by Asmat \textit{et al.}\cite{Asmat:2015}. In an isolated disk, the vortex core is displaced from the center of the disk for the $\bm{X}$ position by application of an in-plane magnetic field and considering a large damping ($\alpha = 1$) for faster convergence; in this position we measured the magnetization $\bm{M}$. Knowing these two quantities, we can now make use of Eq. \ref{lambda}. We repeat the same procedure for each value of K$_{\sigma}$. In the linear regime, the magnetization $\bm{M}$  and $\bm{X}$ are proportional\cite{Sukhostavets:2011}, therefore the $\lambda$ parameter is independent of time and it has an unique value for each K$_{\sigma}$.

\begin{figure}[h]
\centering
\includegraphics[width=1\columnwidth]{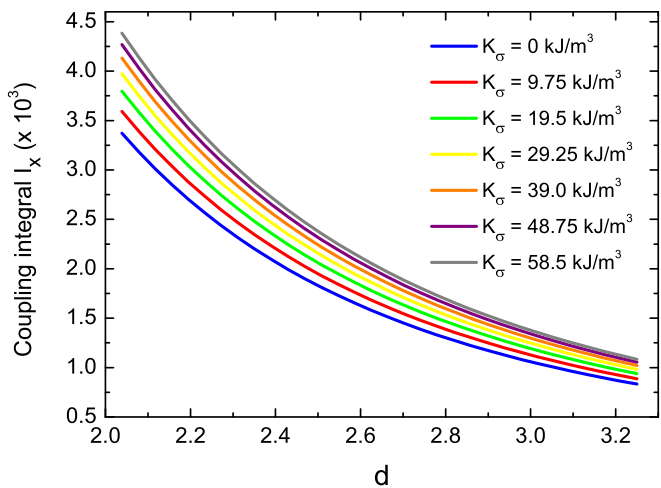}
\caption{Coupling integral I$_x$ as a function of the reduced distance d = D/R for disks with L = 7\:nm and diameter 256\:nm. These results were obtained from Eq. (\ref{integraisdeacoplamento}).}\label{fig:interacciones}
\end{figure}

Accordingly, we find the values of the coupling integrals; these results are shown in Fig. \ref{fig:interacciones}. For d = 2.27 and K$_{\sigma}$ = 0\:kJ/m$^3$ the ratio $\|$Iy$\|$/Ix is approximately 0.38 using expressions obtained by Shibata \textit{et al.}\cite{Shibata:2003}, whereas if we use Eq. \ref{integraisdeacoplamento} the ratio is 0.5. This difference is expected because our model only considers the dipolar term in the interaction energy (Eq. \ref{interacciondipolar1}), however for larger d the ratio $\|$Iy$\|$/Ix begins to approach 0.5.

Considering our expressions for the coupling integrals (Eq. \ref{integraisdeacoplamento}), we will determine the expression for the eigenfrequencies (or coupling frequencies) of the two disks coupled system.

The lagrangian expression for a pair of coupled disks based on the constant of the Thiele's equation is defined by:\cite{Shibata:2006, Asmat:2015}

\begin{eqnarray}\label{lagrangiana}
\mathcal{L} = -\frac{1}{2}\sum_{i}^j\{Gp_i(x_i\dot{y}_i - y_i\dot{x}_i) - \kappa |\bm{X_i|^2}         \}\nonumber \\
- \sum_{i<j}E_{dip}^{ij},
\end{eqnarray}

From the first variation of the lagrangian (Eq. \ref{lagrangiana}) we obtained the equations of motion:

\begin{equation}\label{matriz}
\begin{bmatrix}
\dot{x}_1\\
\dot{x}_2\\
\dot{y}_1\\
\dot{y}_2
\end{bmatrix}
= \frac{1}{G}
\begin{bmatrix}
0 & 0 & p_1\kappa & -2p_1\eta^*\\
0 & 0 & -2p_2\eta^* & p_2\kappa\\
-p_1\kappa & -p_1\eta^* & 0 & 0\\
-p_2\eta^* & -p_2\kappa & 0 & 0
\end{bmatrix}
\begin{bmatrix}
x_1\\
x_2\\
y_1\\
y_2
\end{bmatrix}
\end{equation}

The coupling frequencies (or eigenfrequencies) are obtained from  Eq. \ref{matriz}:

\begin{equation}\label{eigenfrequencies}
\omega_{1,2} = \sqrt{\omega_0^2 - 2p(\frac{\eta^*}{G})^2 \pm \frac{\eta^*}{G}\omega_0\sqrt{5 - 4p}}
\end{equation}

and the splitting frequency:

\begin{equation}\label{splitting_frequency}
|\omega_2 - \omega_1| =  \sqrt{2}\omega_0\sqrt{1-2pu^2 - \sqrt{4u^4 - 5u^2 + 1}}
\end{equation}

where $\omega_0$ is an eigenfrequency of an isolated disk and u = $\eta^*/G\omega_0$.\newline

\indent The coupling frequencies do not depend on the sign of the circulations $C_1$ and $C_2$, they depend only on the combination of polarities (p = +1 or p = -1).\\
\indent The dependence of the coupling integrals on K$_{\sigma}$ and reduced separation distance d is shown in Fig. \ref{fig:interacciones}. The values of the coupling integrals increase with increasing K$_{\sigma}$, meaning that the presence of IPUA favors the appearance of magnetic charges, making the coupling between both disks stronger. This result is very important because as the coupling integrals increase,  the splitting frequency also increases (see Eq. \ref{splitting_frequency}), therefore $\tau$ must decrease (previous Section). This explains why $\tau$ decreases with increasing K$_{\sigma}$, as already discussed in the previous section. The effects caused by the IPUA strongly contrast to those produced by perpendicular magnetic fields. While it is true that in the case of an isolated disk, both the IPUA and the application of a perpendicular magnetic field antiparallel to the polarity of the vortex produce the same effect of decreasing frequency, it is not the same in the case of coupled disks. $\tau$ increases with the increase of the applied  perpendicular magnetic field, while it decreases with increasing K$_{\sigma}$. The reason for this is that the perpendicular magnetic field deforms the core profile of the vortex, making the coupling integrals weak\cite{Han:2012} while the presence of IPUA does not alter the core profile of the vortex\cite{Roy:2013}. 
\indent As expected, the values of the coupling integrals decrease with the increase of the center-to-center distance of the disks.

\begin{figure}[h]
\centering
\includegraphics[width=1\columnwidth]{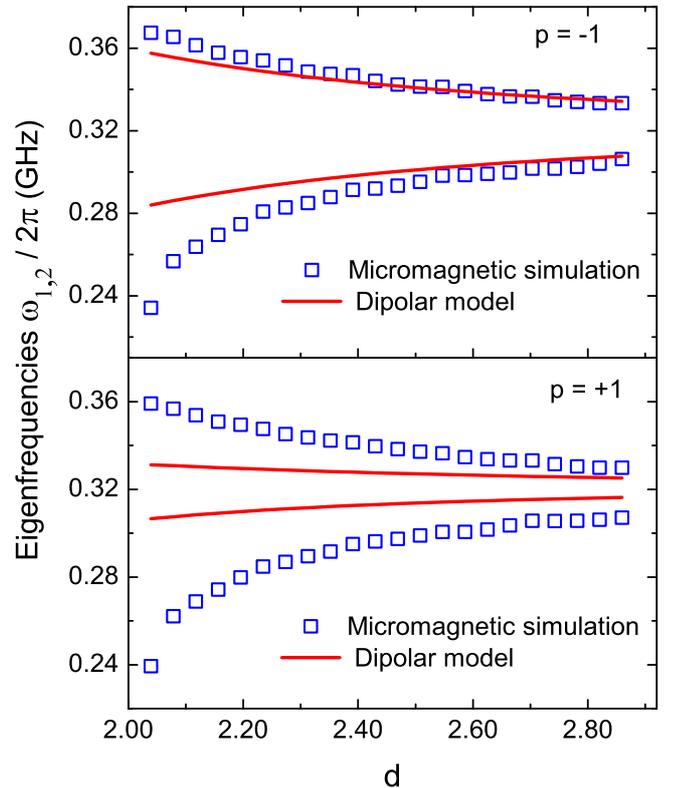}
\caption{Variation of coupling frequencies $\omega_{1,2}/2\pi$ with the reduced distance  d = D/R between two disks with combined polarities p = +1 and p = -1 for K$_{\sigma}$ = 58.5 kJ/m$^3$. The blue squares represent the values obtained from micromagnetic simulations and the red solid line represents the values obtained from Eq. (\ref{eigenfrequencies}).}\label{fig:split}
\end{figure}

In order to test  Eq. \ref{eigenfrequencies}, we compared the coupling frequencies  of the two-coupled disk system obtained from micromagnetic simulations with the results obtained using Eq. \ref{eigenfrequencies}. These results are shown in Fig. (\ref{fig:split}) for K$_{\sigma}$ =  58.5 kJ/m$^3$ and for the combination of polarities p = +1 and p = -1.
As our model considers purely dipolar interaction, which is far-reaching, it is expected that this is well-behaved when the separation distance between the disks is larger, as shown in Fig. \ref{fig:split}.
Although it is necessary to make some adjustments to our model to improve efficiency for small separation distances where high-order magnetic interactions, such as dipole-octupole, octupole-octupole, dipole-triacontadipole are appreciable\cite{Sukhostavets:2013}, it is sufficient to explain the influence of the IPUA in the interaction between two disks.

\section{Conclusion}
In this work we have studied the influence of the IPUA on an isolated disk and on a system of coupled identical disks, both with a magnetic vortex configuration.
In the isolated disk system, we have obtained the gyrotropic frequency,  gyrotropic constant and stiffness constant from micromagnetic simulations. 
In the two coupled disks system, using micromagnetic simulations, we demonstrated that it is possible to control the mutual energy transfer time using the IPUA, introducing a new method for controlling $\tau$. Using a simple analytical dipolar model, we have obtained  the coupling integrals (interactions in the x and y directions) depending on the reduced separation distance d and  K$_{\sigma}$. 
Also, we were able to explain why $\tau$ decreases with increasing K$_{\sigma}$, and clarify the difference in using perpendicular magnetic fields and IPUA.
We have also analytically found the coupling frequencies and compared this with the coupling frequencies obtained from micromagnetic simulations. Our analytical results are consistent with those obtained by micromagnetic simulation.

\section*{ACKNOWLEDGMENTS}
The authors would like to thank the support of the Brazilian agencies CNPq and FAPERJ.

\bibliography{anisotropia}

\begin{thebibliography}{34}%
\makeatletter
\providecommand \@ifxundefined [1]{%
 \@ifx{#1\undefined}
}%
\providecommand \@ifnum [1]{%
 \ifnum #1\expandafter \@firstoftwo
 \else \expandafter \@secondoftwo
 \fi
}%
\providecommand \@ifx [1]{%
 \ifx #1\expandafter \@firstoftwo
 \else \expandafter \@secondoftwo
 \fi
}%
\providecommand \natexlab [1]{#1}%
\providecommand \enquote  [1]{``#1''}%
\providecommand \bibnamefont  [1]{#1}%
\providecommand \bibfnamefont [1]{#1}%
\providecommand \citenamefont [1]{#1}%
\providecommand \href@noop [0]{\@secondoftwo}%
\providecommand \href [0]{\begingroup \@sanitize@url \@href}%
\providecommand \@href[1]{\@@startlink{#1}\@@href}%
\providecommand \@@href[1]{\endgroup#1\@@endlink}%
\providecommand \@sanitize@url [0]{\catcode `\\12\catcode `\$12\catcode
  `\&12\catcode `\#12\catcode `\^12\catcode `\_12\catcode `\%12\relax}%
\providecommand \@@startlink[1]{}%
\providecommand \@@endlink[0]{}%
\providecommand \url  [0]{\begingroup\@sanitize@url \@url }%
\providecommand \@url [1]{\endgroup\@href {#1}{\urlprefix }}%
\providecommand \urlprefix  [0]{URL }%
\providecommand \Eprint [0]{\href }%
\providecommand \doibase [0]{http://dx.doi.org/}%
\providecommand \selectlanguage [0]{\@gobble}%
\providecommand \bibinfo  [0]{\@secondoftwo}%
\providecommand \bibfield  [0]{\@secondoftwo}%
\providecommand \translation [1]{[#1]}%
\providecommand \BibitemOpen [0]{}%
\providecommand \bibitemStop [0]{}%
\providecommand \bibitemNoStop [0]{.\EOS\space}%
\providecommand \EOS [0]{\spacefactor3000\relax}%
\providecommand \BibitemShut  [1]{\csname bibitem#1\endcsname}%
\let\auto@bib@innerbib\@empty
\bibitem [{\citenamefont {Guslienko}(2008)}]{Guslienko:2008}%
  \BibitemOpen
  \bibfield  {author} {\bibinfo {author} {\bibfnamefont {K.~Y.}\ \bibnamefont
  {Guslienko}},\ }\href {\doibase doi:10.1166/jnn.2008.003} {\bibfield
  {journal} {\bibinfo  {journal} {J. Nanoscience Nanotechnol.}\ }\textbf
  {\bibinfo {volume} {8}},\ \bibinfo {pages} {2745} (\bibinfo {year}
  {2008})}\BibitemShut {NoStop}%
\bibitem [{\citenamefont {Guimar{\~a}es}(2009)}]{Guimaraes:2009}%
  \BibitemOpen
  \bibfield  {author} {\bibinfo {author} {\bibfnamefont {A.~P.}\ \bibnamefont
  {Guimar{\~a}es}},\ }\href@noop {} {\emph {\bibinfo {title} {Principles of
  Nanomagnetism}}}\ (\bibinfo  {publisher} {Springer},\ \bibinfo {address}
  {Berlin},\ \bibinfo {year} {2009})\BibitemShut {NoStop}%
\bibitem [{\citenamefont {Locatelli}\ \emph {et~al.}(2014)\citenamefont
  {Locatelli}, \citenamefont {Cros},\ and\ \citenamefont
  {Grollier}}]{Locatelli:2014}%
  \BibitemOpen
  \bibfield  {author} {\bibinfo {author} {\bibfnamefont {N.}~\bibnamefont
  {Locatelli}}, \bibinfo {author} {\bibfnamefont {V.}~\bibnamefont {Cros}}, \
  and\ \bibinfo {author} {\bibfnamefont {J.}~\bibnamefont {Grollier}},\ }\href
  {\doibase http://dx.doi.org/10.1038/nmat3823} {\bibfield  {journal} {\bibinfo
   {journal} {Nat. Mater.}\ }\textbf {\bibinfo {volume} {13}},\ \bibinfo
  {pages} {1476} (\bibinfo {year} {2014})}\BibitemShut {NoStop}%
\bibitem [{\citenamefont {Bohlens}\ \emph {et~al.}(2008)\citenamefont
  {Bohlens}, \citenamefont {Kr{\"u}ger}, \citenamefont {Drews}, \citenamefont
  {Bolte}, \citenamefont {Meier},\ and\ \citenamefont
  {Pfannkuche}}]{Bohlens:2008}%
  \BibitemOpen
  \bibfield  {author} {\bibinfo {author} {\bibfnamefont {S.}~\bibnamefont
  {Bohlens}}, \bibinfo {author} {\bibfnamefont {B.}~\bibnamefont {Kr{\"u}ger}},
  \bibinfo {author} {\bibfnamefont {A.}~\bibnamefont {Drews}}, \bibinfo
  {author} {\bibfnamefont {M.}~\bibnamefont {Bolte}}, \bibinfo {author}
  {\bibfnamefont {G.}~\bibnamefont {Meier}}, \ and\ \bibinfo {author}
  {\bibfnamefont {D.}~\bibnamefont {Pfannkuche}},\ }\href {\doibase
  http://dx.doi.org/10.1063/1.2998584} {\bibfield  {journal} {\bibinfo
  {journal} {Appl. Phys. Lett.}\ }\textbf {\bibinfo {volume} {93}},\ \bibinfo
  {pages} {142508} (\bibinfo {year} {2008})}\BibitemShut {NoStop}%
\bibitem [{\citenamefont {Belanovsky}\ \emph {et~al.}(2013)\citenamefont
  {Belanovsky}, \citenamefont {Locatelli}, \citenamefont {Skirdkov},
  \citenamefont {Abreu~Araujo}, \citenamefont {Zvezdin}, \citenamefont
  {Grollier}, \citenamefont {Cros},\ and\ \citenamefont
  {Zvezdin}}]{Belanovsky:2013}%
  \BibitemOpen
  \bibfield  {author} {\bibinfo {author} {\bibfnamefont {A.~D.}\ \bibnamefont
  {Belanovsky}}, \bibinfo {author} {\bibfnamefont {N.}~\bibnamefont
  {Locatelli}}, \bibinfo {author} {\bibfnamefont {P.~N.}\ \bibnamefont
  {Skirdkov}}, \bibinfo {author} {\bibfnamefont {F.}~\bibnamefont
  {Abreu~Araujo}}, \bibinfo {author} {\bibfnamefont {K.~A.}\ \bibnamefont
  {Zvezdin}}, \bibinfo {author} {\bibfnamefont {J.}~\bibnamefont {Grollier}},
  \bibinfo {author} {\bibfnamefont {V.}~\bibnamefont {Cros}}, \ and\ \bibinfo
  {author} {\bibfnamefont {A.~K.}\ \bibnamefont {Zvezdin}},\ }\href {\doibase
  http://dx.doi.org/10.1063/1.4821073} {\bibfield  {journal} {\bibinfo
  {journal} {Appl. Phys. Lett.}\ }\textbf {\bibinfo {volume} {103}},\ \bibinfo
  {eid} {122405} (\bibinfo {year} {2013})}\BibitemShut {NoStop}%
\bibitem [{\citenamefont {Novosad}\ \emph {et~al.}(2005)\citenamefont
  {Novosad}, \citenamefont {Fradin}, \citenamefont {Roy}, \citenamefont
  {Buchanan}, \citenamefont {Guslienko},\ and\ \citenamefont
  {Bader}}]{Novosad:2005}%
  \BibitemOpen
  \bibfield  {author} {\bibinfo {author} {\bibfnamefont {V.}~\bibnamefont
  {Novosad}}, \bibinfo {author} {\bibfnamefont {F.~Y.}\ \bibnamefont {Fradin}},
  \bibinfo {author} {\bibfnamefont {P.~E.}\ \bibnamefont {Roy}}, \bibinfo
  {author} {\bibfnamefont {K.~S.}\ \bibnamefont {Buchanan}}, \bibinfo {author}
  {\bibfnamefont {K.~Y.}\ \bibnamefont {Guslienko}}, \ and\ \bibinfo {author}
  {\bibfnamefont {S.~D.}\ \bibnamefont {Bader}},\ }\href {\doibase
  10.1103/PhysRevB.72.024455} {\bibfield  {journal} {\bibinfo  {journal} {Phys.
  Rev. B}\ }\textbf {\bibinfo {volume} {72}},\ \bibinfo {pages} {024455}
  (\bibinfo {year} {2005})}\BibitemShut {NoStop}%
\bibitem [{\citenamefont {Guslienko}\ \emph {et~al.}(2006)\citenamefont
  {Guslienko}, \citenamefont {Han}, \citenamefont {Keavney}, \citenamefont
  {Divan},\ and\ \citenamefont {Bader}}]{Guslienko:2006}%
  \BibitemOpen
  \bibfield  {author} {\bibinfo {author} {\bibfnamefont {K.~Y.}\ \bibnamefont
  {Guslienko}}, \bibinfo {author} {\bibfnamefont {X.~F.}\ \bibnamefont {Han}},
  \bibinfo {author} {\bibfnamefont {D.~J.}\ \bibnamefont {Keavney}}, \bibinfo
  {author} {\bibfnamefont {R.}~\bibnamefont {Divan}}, \ and\ \bibinfo {author}
  {\bibfnamefont {S.~D.}\ \bibnamefont {Bader}},\ }\href {\doibase
  10.1103/PhysRevLett.96.067205} {\bibfield  {journal} {\bibinfo  {journal}
  {Phys. Rev. Lett.}\ }\textbf {\bibinfo {volume} {96}},\ \bibinfo {pages}
  {067205} (\bibinfo {year} {2006})}\BibitemShut {NoStop}%
\bibitem [{\citenamefont {Novosad}\ \emph {et~al.}(2002)\citenamefont
  {Novosad}, \citenamefont {Grimsditch}, \citenamefont {Guslienko},
  \citenamefont {Vavassori}, \citenamefont {Otani},\ and\ \citenamefont
  {Bader}}]{Novosad:2002}%
  \BibitemOpen
  \bibfield  {author} {\bibinfo {author} {\bibfnamefont {V.}~\bibnamefont
  {Novosad}}, \bibinfo {author} {\bibfnamefont {M.}~\bibnamefont {Grimsditch}},
  \bibinfo {author} {\bibfnamefont {K.~Y.}\ \bibnamefont {Guslienko}}, \bibinfo
  {author} {\bibfnamefont {P.}~\bibnamefont {Vavassori}}, \bibinfo {author}
  {\bibfnamefont {Y.}~\bibnamefont {Otani}}, \ and\ \bibinfo {author}
  {\bibfnamefont {S.~D.}\ \bibnamefont {Bader}},\ }\href {\doibase
  10.1103/PhysRevB.66.052407} {\bibfield  {journal} {\bibinfo  {journal} {Phys.
  Rev. B}\ }\textbf {\bibinfo {volume} {66}},\ \bibinfo {pages} {052407}
  (\bibinfo {year} {2002})}\BibitemShut {NoStop}%
\bibitem [{\citenamefont {Yoo}\ \emph {et~al.}(2011)\citenamefont {Yoo},
  \citenamefont {Lee}, \citenamefont {Han},\ and\ \citenamefont
  {Kim}}]{Yoo:2011}%
  \BibitemOpen
  \bibfield  {author} {\bibinfo {author} {\bibfnamefont {M.-W.}\ \bibnamefont
  {Yoo}}, \bibinfo {author} {\bibfnamefont {K.-S.}\ \bibnamefont {Lee}},
  \bibinfo {author} {\bibfnamefont {D.-S.}\ \bibnamefont {Han}}, \ and\
  \bibinfo {author} {\bibfnamefont {S.-K.}\ \bibnamefont {Kim}},\ }\href
  {\doibase http://dx.doi.org/10.1063/1.3563561} {\bibfield  {journal}
  {\bibinfo  {journal} {J. Appl. Phys.}\ }\textbf {\bibinfo {volume} {109}},\
  \bibinfo {pages} {063903} (\bibinfo {year} {2011})}\BibitemShut {NoStop}%
\bibitem [{\citenamefont {Choi}\ \emph {et~al.}(2008)\citenamefont {Choi},
  \citenamefont {Kim}, \citenamefont {Lee},\ and\ \citenamefont
  {Yu}}]{Choi:2008}%
  \BibitemOpen
  \bibfield  {author} {\bibinfo {author} {\bibfnamefont {Y.-S.}\ \bibnamefont
  {Choi}}, \bibinfo {author} {\bibfnamefont {S.-K.}\ \bibnamefont {Kim}},
  \bibinfo {author} {\bibfnamefont {K.-S.}\ \bibnamefont {Lee}}, \ and\
  \bibinfo {author} {\bibfnamefont {Y.-S.}\ \bibnamefont {Yu}},\ }\href
  {\doibase http://dx.doi.org/10.1063/1.3012380} {\bibfield  {journal}
  {\bibinfo  {journal} {Appl. Phys. Lett}\ }\textbf {\bibinfo {volume} {93}},\
  \bibinfo {pages} {182508} (\bibinfo {year} {2008})}\BibitemShut {NoStop}%
\bibitem [{\citenamefont {de~Loubens}\ \emph {et~al.}(2009)\citenamefont
  {de~Loubens}, \citenamefont {Riegler}, \citenamefont {Pigeau}, \citenamefont
  {Lochner}, \citenamefont {Boust}, \citenamefont {Guslienko}, \citenamefont
  {Hurdequint}, \citenamefont {Molenkamp}, \citenamefont {Schmidt},
  \citenamefont {Slavin}, \citenamefont {Tiberkevich}, \citenamefont
  {Vukadinovic},\ and\ \citenamefont {Klein}}]{Loubens:2009}%
  \BibitemOpen
  \bibfield  {author} {\bibinfo {author} {\bibfnamefont {G.}~\bibnamefont
  {de~Loubens}}, \bibinfo {author} {\bibfnamefont {A.}~\bibnamefont {Riegler}},
  \bibinfo {author} {\bibfnamefont {B.}~\bibnamefont {Pigeau}}, \bibinfo
  {author} {\bibfnamefont {F.}~\bibnamefont {Lochner}}, \bibinfo {author}
  {\bibfnamefont {F.}~\bibnamefont {Boust}}, \bibinfo {author} {\bibfnamefont
  {K.~Y.}\ \bibnamefont {Guslienko}}, \bibinfo {author} {\bibfnamefont
  {H.}~\bibnamefont {Hurdequint}}, \bibinfo {author} {\bibfnamefont {L.~W.}\
  \bibnamefont {Molenkamp}}, \bibinfo {author} {\bibfnamefont {G.}~\bibnamefont
  {Schmidt}}, \bibinfo {author} {\bibfnamefont {A.~N.}\ \bibnamefont {Slavin}},
  \bibinfo {author} {\bibfnamefont {V.~S.}\ \bibnamefont {Tiberkevich}},
  \bibinfo {author} {\bibfnamefont {N.}~\bibnamefont {Vukadinovic}}, \ and\
  \bibinfo {author} {\bibfnamefont {O.}~\bibnamefont {Klein}},\ }\href
  {\doibase 10.1103/PhysRevLett.102.177602} {\bibfield  {journal} {\bibinfo
  {journal} {Phys. Rev. Lett.}\ }\textbf {\bibinfo {volume} {102}},\ \bibinfo
  {pages} {177602} (\bibinfo {year} {2009})}\BibitemShut {NoStop}%
\bibitem [{\citenamefont {Cavill}\ \emph {et~al.}(2013)\citenamefont {Cavill},
  \citenamefont {Parkes}, \citenamefont {Miguel}, \citenamefont {Dhesi},
  \citenamefont {Edmonds}, \citenamefont {Campion},\ and\ \citenamefont
  {Rushforth}}]{Cavill:2013}%
  \BibitemOpen
  \bibfield  {author} {\bibinfo {author} {\bibfnamefont {S.~A.}\ \bibnamefont
  {Cavill}}, \bibinfo {author} {\bibfnamefont {D.~E.}\ \bibnamefont {Parkes}},
  \bibinfo {author} {\bibfnamefont {J.}~\bibnamefont {Miguel}}, \bibinfo
  {author} {\bibfnamefont {S.~S.}\ \bibnamefont {Dhesi}}, \bibinfo {author}
  {\bibfnamefont {K.~W.}\ \bibnamefont {Edmonds}}, \bibinfo {author}
  {\bibfnamefont {R.~P.}\ \bibnamefont {Campion}}, \ and\ \bibinfo {author}
  {\bibfnamefont {A.~W.}\ \bibnamefont {Rushforth}},\ }\href {\doibase
  http://dx.doi.org/10.1063/1.4789396} {\bibfield  {journal} {\bibinfo
  {journal} {Appl. Phys. Lett.}\ }\textbf {\bibinfo {volume} {102}} (\bibinfo
  {year} {2013}),\ http://dx.doi.org/10.1063/1.4789396}\BibitemShut {NoStop}%
\bibitem [{\citenamefont {Roy}(2013)}]{Roy:2013}%
  \BibitemOpen
  \bibfield  {author} {\bibinfo {author} {\bibfnamefont {P.~E.}\ \bibnamefont
  {Roy}},\ }\href {\doibase http://dx.doi.org/10.1063/1.4802976} {\bibfield
  {journal} {\bibinfo  {journal} {Appl. Phys. Lett.}\ }\textbf {\bibinfo
  {volume} {102}},\ \bibinfo {pages} {162411} (\bibinfo {year}
  {2013})}\BibitemShut {NoStop}%
\bibitem [{\citenamefont {Ostler}\ \emph {et~al.}(2015)\citenamefont {Ostler},
  \citenamefont {Cuadrado}, \citenamefont {Chantrell}, \citenamefont
  {Rushforth},\ and\ \citenamefont {Cavill}}]{Ostler:2015}%
  \BibitemOpen
  \bibfield  {author} {\bibinfo {author} {\bibfnamefont {T.~A.}\ \bibnamefont
  {Ostler}}, \bibinfo {author} {\bibfnamefont {R.}~\bibnamefont {Cuadrado}},
  \bibinfo {author} {\bibfnamefont {R.~W.}\ \bibnamefont {Chantrell}}, \bibinfo
  {author} {\bibfnamefont {A.~W.}\ \bibnamefont {Rushforth}}, \ and\ \bibinfo
  {author} {\bibfnamefont {S.~A.}\ \bibnamefont {Cavill}},\ }\href {\doibase
  10.1103/PhysRevLett.115.067202} {\bibfield  {journal} {\bibinfo  {journal}
  {Phys. Rev. Lett.}\ }\textbf {\bibinfo {volume} {115}},\ \bibinfo {pages}
  {067202} (\bibinfo {year} {2015})}\BibitemShut {NoStop}%
\bibitem [{\citenamefont {Parreiras}\ and\ \citenamefont
  {Martins}(2015)}]{Parreiras:2015}%
  \BibitemOpen
  \bibfield  {author} {\bibinfo {author} {\bibfnamefont {S.}~\bibnamefont
  {Parreiras}}\ and\ \bibinfo {author} {\bibfnamefont {M.}~\bibnamefont
  {Martins}},\ }\href {\doibase 10.1016/j.phpro.2015.12.185} {\bibfield
  {journal} {\bibinfo  {journal} {Phys. Proc.}\ }\textbf {\bibinfo {volume}
  {75}},\ \bibinfo {pages} {1142 } (\bibinfo {year} {2015})}\BibitemShut
  {NoStop}%
\bibitem [{\citenamefont {Novais}\ \emph {et~al.}(2011)\citenamefont {Novais},
  \citenamefont {Landeros}, \citenamefont {Barbosa}, \citenamefont {Martins},
  \citenamefont {Garcia},\ and\ \citenamefont {Guimar{\~a}es}}]{Novais:2011}%
  \BibitemOpen
  \bibfield  {author} {\bibinfo {author} {\bibfnamefont {E.~R.~P.}\
  \bibnamefont {Novais}}, \bibinfo {author} {\bibfnamefont {P.}~\bibnamefont
  {Landeros}}, \bibinfo {author} {\bibfnamefont {A.~G.~S.}\ \bibnamefont
  {Barbosa}}, \bibinfo {author} {\bibfnamefont {M.~D.}\ \bibnamefont
  {Martins}}, \bibinfo {author} {\bibfnamefont {F.}~\bibnamefont {Garcia}}, \
  and\ \bibinfo {author} {\bibfnamefont {A.~P.}\ \bibnamefont
  {Guimar{\~a}es}},\ }\href {\doibase http://dx.doi.org/10.1063/1.3631081}
  {\bibfield  {journal} {\bibinfo  {journal} {J. Appl. Phys.}\ }\textbf
  {\bibinfo {volume} {110}},\ \bibinfo {pages} {053917} (\bibinfo {year}
  {2011})}\BibitemShut {NoStop}%
\bibitem [{\citenamefont {Novais}\ \emph {et~al.}(2013)\citenamefont {Novais},
  \citenamefont {Allende}, \citenamefont {Altbir}, \citenamefont {Landeros},
  \citenamefont {Garcia},\ and\ \citenamefont {Guimar{\~a}es}}]{Novais:2013}%
  \BibitemOpen
  \bibfield  {author} {\bibinfo {author} {\bibfnamefont {E.~R.~P.}\
  \bibnamefont {Novais}}, \bibinfo {author} {\bibfnamefont {S.}~\bibnamefont
  {Allende}}, \bibinfo {author} {\bibfnamefont {D.}~\bibnamefont {Altbir}},
  \bibinfo {author} {\bibfnamefont {P.}~\bibnamefont {Landeros}}, \bibinfo
  {author} {\bibfnamefont {F.}~\bibnamefont {Garcia}}, \ and\ \bibinfo {author}
  {\bibfnamefont {A.~P.}\ \bibnamefont {Guimar{\~a}es}},\ }\href {\doibase
  http://dx.doi.org/10.1063/1.4824803} {\bibfield  {journal} {\bibinfo
  {journal} {J. Appl. Phys.}\ }\textbf {\bibinfo {volume} {114}},\ \bibinfo
  {eid} {153905} (\bibinfo {year} {2013})}\BibitemShut {NoStop}%
\bibitem [{\citenamefont {Garcia}\ \emph {et~al.}(2010)\citenamefont {Garcia},
  \citenamefont {Westfahl}, \citenamefont {Schoenmaker}, \citenamefont
  {Carvalho}, \citenamefont {Santos}, \citenamefont {Pojar}, \citenamefont
  {Seabra}, \citenamefont {Belkhou}, \citenamefont {Bendounan}, \citenamefont
  {Novais},\ and\ \citenamefont {Guimar{\~a}es}}]{Garcia:2010}%
  \BibitemOpen
  \bibfield  {author} {\bibinfo {author} {\bibfnamefont {F.}~\bibnamefont
  {Garcia}}, \bibinfo {author} {\bibfnamefont {H.}~\bibnamefont {Westfahl}},
  \bibinfo {author} {\bibfnamefont {J.}~\bibnamefont {Schoenmaker}}, \bibinfo
  {author} {\bibfnamefont {E.~J.}\ \bibnamefont {Carvalho}}, \bibinfo {author}
  {\bibfnamefont {A.~D.}\ \bibnamefont {Santos}}, \bibinfo {author}
  {\bibfnamefont {M.}~\bibnamefont {Pojar}}, \bibinfo {author} {\bibfnamefont
  {A.~C.}\ \bibnamefont {Seabra}}, \bibinfo {author} {\bibfnamefont
  {R.}~\bibnamefont {Belkhou}}, \bibinfo {author} {\bibfnamefont
  {A.}~\bibnamefont {Bendounan}}, \bibinfo {author} {\bibfnamefont {E.~R.~P.}\
  \bibnamefont {Novais}}, \ and\ \bibinfo {author} {\bibfnamefont {A.~P.}\
  \bibnamefont {Guimar{\~a}es}},\ }\href {\doibase
  http://dx.doi.org/10.1063/1.3462305} {\bibfield  {journal} {\bibinfo
  {journal} {Appl. Phys. Lett.}\ }\textbf {\bibinfo {volume} {97}},\ \bibinfo
  {pages} {022501} (\bibinfo {year} {2010})}\BibitemShut {NoStop}%
\bibitem [{\citenamefont {Fior}\ \emph {et~al.}(2016)\citenamefont {Fior},
  \citenamefont {Novais}, \citenamefont {Sinnecker}, \citenamefont
  {Guimar{\~a}es},\ and\ \citenamefont {Garcia}}]{Fior:2016}%
  \BibitemOpen
  \bibfield  {author} {\bibinfo {author} {\bibfnamefont {G.~B.~M.}\
  \bibnamefont {Fior}}, \bibinfo {author} {\bibfnamefont {E.~R.~P.}\
  \bibnamefont {Novais}}, \bibinfo {author} {\bibfnamefont {J.~P.}\
  \bibnamefont {Sinnecker}}, \bibinfo {author} {\bibfnamefont {A.~P.}\
  \bibnamefont {Guimar{\~a}es}}, \ and\ \bibinfo {author} {\bibfnamefont
  {F.}~\bibnamefont {Garcia}},\ }\href {\doibase
  http://dx.doi.org/10.1063/1.4942534} {\bibfield  {journal} {\bibinfo
  {journal} {J. Appl. Phys.}\ }\textbf {\bibinfo {volume} {119}},\ \bibinfo
  {pages} {093906} (\bibinfo {year} {2016})}\BibitemShut {NoStop}%
\bibitem [{\citenamefont {Parkes}\ \emph {et~al.}(2013)\citenamefont {Parkes},
  \citenamefont {Shelford}, \citenamefont {Wadley}, \citenamefont {Holý},
  \citenamefont {Wang}, \citenamefont {Hindmarch}, \citenamefont {van~der
  Laan}, \citenamefont {Campion}, \citenamefont {Edmonds}, \citenamefont
  {Cavill},\ and\ \citenamefont {Rushforth}}]{Parkes:2013}%
  \BibitemOpen
  \bibfield  {author} {\bibinfo {author} {\bibfnamefont {D.~E.}\ \bibnamefont
  {Parkes}}, \bibinfo {author} {\bibfnamefont {L.~R.}\ \bibnamefont
  {Shelford}}, \bibinfo {author} {\bibfnamefont {P.}~\bibnamefont {Wadley}},
  \bibinfo {author} {\bibfnamefont {V.}~\bibnamefont {Holý}}, \bibinfo
  {author} {\bibfnamefont {M.}~\bibnamefont {Wang}}, \bibinfo {author}
  {\bibfnamefont {A.~T.}\ \bibnamefont {Hindmarch}}, \bibinfo {author}
  {\bibfnamefont {G.}~\bibnamefont {van~der Laan}}, \bibinfo {author}
  {\bibfnamefont {R.~P.}\ \bibnamefont {Campion}}, \bibinfo {author}
  {\bibfnamefont {K.~W.}\ \bibnamefont {Edmonds}}, \bibinfo {author}
  {\bibfnamefont {S.~A.}\ \bibnamefont {Cavill}}, \ and\ \bibinfo {author}
  {\bibfnamefont {A.~W.}\ \bibnamefont {Rushforth}},\ }\href {\doibase
  http://dx.doi.org/10.1038/srep02220} {\bibfield  {journal} {\bibinfo
  {journal} {Sci. Rep.}\ }\textbf {\bibinfo {volume} {3}},\ \bibinfo {pages}
  {2220} (\bibinfo {year} {2013})}\BibitemShut {NoStop}%
\bibitem [{\citenamefont {Shibata}\ \emph {et~al.}(2003)\citenamefont
  {Shibata}, \citenamefont {Shigeto},\ and\ \citenamefont
  {Otani}}]{Shibata:2003}%
  \BibitemOpen
  \bibfield  {author} {\bibinfo {author} {\bibfnamefont {J.}~\bibnamefont
  {Shibata}}, \bibinfo {author} {\bibfnamefont {K.}~\bibnamefont {Shigeto}}, \
  and\ \bibinfo {author} {\bibfnamefont {Y.}~\bibnamefont {Otani}},\ }\href
  {\doibase 10.1103/PhysRevB.67.224404} {\bibfield  {journal} {\bibinfo
  {journal} {Phys. Rev. B}\ }\textbf {\bibinfo {volume} {67}},\ \bibinfo
  {pages} {224404} (\bibinfo {year} {2003})}\BibitemShut {NoStop}%
\bibitem [{\citenamefont {Sukhostavets}\ \emph {et~al.}(2011)\citenamefont
  {Sukhostavets}, \citenamefont {Gonzalez},\ and\ \citenamefont
  {Guslienko}}]{Sukhostavets:2011}%
  \BibitemOpen
  \bibfield  {author} {\bibinfo {author} {\bibfnamefont {O.~V.}\ \bibnamefont
  {Sukhostavets}}, \bibinfo {author} {\bibfnamefont {J.~M.}\ \bibnamefont
  {Gonzalez}}, \ and\ \bibinfo {author} {\bibfnamefont {K.~Y.}\ \bibnamefont
  {Guslienko}},\ }\href {http://stacks.iop.org/1882-0786/4/i=6/a=065003}
  {\bibfield  {journal} {\bibinfo  {journal} {Appl. Phys. Express}\ }\textbf
  {\bibinfo {volume} {4}},\ \bibinfo {pages} {065003} (\bibinfo {year}
  {2011})}\BibitemShut {NoStop}%
\bibitem [{\citenamefont {Sinnecker}\ \emph {et~al.}(2014)\citenamefont
  {Sinnecker}, \citenamefont {Vigo-Cotrina}, \citenamefont {Garcia},
  \citenamefont {Novais},\ and\ \citenamefont
  {Guimar{\~a}es}}]{Sinnecker:2014}%
  \BibitemOpen
  \bibfield  {author} {\bibinfo {author} {\bibfnamefont {J.~P.}\ \bibnamefont
  {Sinnecker}}, \bibinfo {author} {\bibfnamefont {H.}~\bibnamefont
  {Vigo-Cotrina}}, \bibinfo {author} {\bibfnamefont {F.}~\bibnamefont
  {Garcia}}, \bibinfo {author} {\bibfnamefont {E.~R.~P.}\ \bibnamefont
  {Novais}}, \ and\ \bibinfo {author} {\bibfnamefont {A.~P.}\ \bibnamefont
  {Guimar{\~a}es}},\ }\href {\doibase http://dx.doi.org/10.1063/1.4878875}
  {\bibfield  {journal} {\bibinfo  {journal} {J. Appl. Phys.}\ }\textbf
  {\bibinfo {volume} {115}},\ \bibinfo {pages} {203902} (\bibinfo {year}
  {2014})}\BibitemShut {NoStop}%
\bibitem [{\citenamefont {Jung}\ \emph {et~al.}(2011)\citenamefont {Jung},
  \citenamefont {Lee}, \citenamefont {Jeong}, \citenamefont {Choi},
  \citenamefont {Yu}, \citenamefont {Han}, \citenamefont {Vogel}, \citenamefont
  {Bocklage}, \citenamefont {Meier}, \citenamefont {Im}, \citenamefont
  {Fischer},\ and\ \citenamefont {Kim}}]{Jung:2011}%
  \BibitemOpen
  \bibfield  {author} {\bibinfo {author} {\bibfnamefont {H.}~\bibnamefont
  {Jung}}, \bibinfo {author} {\bibfnamefont {K.-S.}\ \bibnamefont {Lee}},
  \bibinfo {author} {\bibfnamefont {D.-E.}\ \bibnamefont {Jeong}}, \bibinfo
  {author} {\bibfnamefont {Y.-S.}\ \bibnamefont {Choi}}, \bibinfo {author}
  {\bibfnamefont {Y.-S.}\ \bibnamefont {Yu}}, \bibinfo {author} {\bibfnamefont
  {D.-S.}\ \bibnamefont {Han}}, \bibinfo {author} {\bibfnamefont
  {A.}~\bibnamefont {Vogel}}, \bibinfo {author} {\bibfnamefont
  {L.}~\bibnamefont {Bocklage}}, \bibinfo {author} {\bibfnamefont
  {G.}~\bibnamefont {Meier}}, \bibinfo {author} {\bibfnamefont {M.-Y.}\
  \bibnamefont {Im}}, \bibinfo {author} {\bibfnamefont {P.}~\bibnamefont
  {Fischer}}, \ and\ \bibinfo {author} {\bibfnamefont {S.-K.}\ \bibnamefont
  {Kim}},\ }\href {\doibase http://dx.doi.org/10.1038/srep00059} {\bibfield
  {journal} {\bibinfo  {journal} {Sci. Rep.}\ }\textbf {\bibinfo {volume}
  {1}},\ \bibinfo {pages} {59} (\bibinfo {year} {2011})}\BibitemShut {NoStop}%
\bibitem [{\citenamefont {Kim}\ \emph {et~al.}(2012)\citenamefont {Kim},
  \citenamefont {Lee}, \citenamefont {Jung}, \citenamefont {Han},\ and\
  \citenamefont {Kim}}]{kim:2012}%
  \BibitemOpen
  \bibfield  {author} {\bibinfo {author} {\bibfnamefont {J.-H.}\ \bibnamefont
  {Kim}}, \bibinfo {author} {\bibfnamefont {K.-S.}\ \bibnamefont {Lee}},
  \bibinfo {author} {\bibfnamefont {H.}~\bibnamefont {Jung}}, \bibinfo {author}
  {\bibfnamefont {D.-S.}\ \bibnamefont {Han}}, \ and\ \bibinfo {author}
  {\bibfnamefont {S.-K.}\ \bibnamefont {Kim}},\ }\href {\doibase
  http://dx.doi.org/10.1063/1.4748885} {\bibfield  {journal} {\bibinfo
  {journal} {Appl. Phys. Lett.}\ }\textbf {\bibinfo {volume} {101}},\ \bibinfo
  {pages} {092403} (\bibinfo {year} {2012})}\BibitemShut {NoStop}%
\bibitem [{\citenamefont {Han}\ \emph {et~al.}(2015)\citenamefont {Han},
  \citenamefont {Cho}, \citenamefont {Jeong},\ and\ \citenamefont
  {Kim}}]{Han:2012}%
  \BibitemOpen
  \bibfield  {author} {\bibinfo {author} {\bibfnamefont {D.-S.}\ \bibnamefont
  {Han}}, \bibinfo {author} {\bibfnamefont {Y.-J.}\ \bibnamefont {Cho}},
  \bibinfo {author} {\bibfnamefont {H.-B.}\ \bibnamefont {Jeong}}, \ and\
  \bibinfo {author} {\bibfnamefont {S.-K.}\ \bibnamefont {Kim}},\ }\href
  {\doibase http://dx.doi.org/10.1063/1.4913503} {\bibfield  {journal}
  {\bibinfo  {journal} {J. Appl. Phys.}\ }\textbf {\bibinfo {volume} {117}},\
  \bibinfo {pages} {083910} (\bibinfo {year} {2015})}\BibitemShut {NoStop}%
\bibitem [{\citenamefont {Vansteenkiste}\ \emph {et~al.}(2014)\citenamefont
  {Vansteenkiste}, \citenamefont {Leliaert}, \citenamefont {Dvornik},
  \citenamefont {Helsen}, \citenamefont {Garcia-Sanchez},\ and\ \citenamefont
  {Van~Waeyenberge}}]{Vansteenkiste:2014}%
  \BibitemOpen
  \bibfield  {author} {\bibinfo {author} {\bibfnamefont {A.}~\bibnamefont
  {Vansteenkiste}}, \bibinfo {author} {\bibfnamefont {J.}~\bibnamefont
  {Leliaert}}, \bibinfo {author} {\bibfnamefont {M.}~\bibnamefont {Dvornik}},
  \bibinfo {author} {\bibfnamefont {M.}~\bibnamefont {Helsen}}, \bibinfo
  {author} {\bibfnamefont {F.}~\bibnamefont {Garcia-Sanchez}}, \ and\ \bibinfo
  {author} {\bibfnamefont {B.}~\bibnamefont {Van~Waeyenberge}},\ }\href
  {\doibase http://dx.doi.org/10.1063/1.4899186} {\bibfield  {journal}
  {\bibinfo  {journal} {AIP Advances}\ }\textbf {\bibinfo {volume} {4}},\
  \bibinfo {pages} {107133} (\bibinfo {year} {2014})}\BibitemShut {NoStop}%
\bibitem [{\citenamefont {Brintlinger}\ \emph {et~al.}(2010)\citenamefont
  {Brintlinger}, \citenamefont {Lim}, \citenamefont {Baloch}, \citenamefont
  {Alexander}, \citenamefont {Qi}, \citenamefont {Barry}, \citenamefont
  {Melngailis}, \citenamefont {Salamanca-Riba}, \citenamefont {Takeuchi},\ and\
  \citenamefont {Cumings}}]{SungHwan:2010}%
  \BibitemOpen
  \bibfield  {author} {\bibinfo {author} {\bibfnamefont {T.}~\bibnamefont
  {Brintlinger}}, \bibinfo {author} {\bibfnamefont {S.-H.}\ \bibnamefont
  {Lim}}, \bibinfo {author} {\bibfnamefont {K.~H.}\ \bibnamefont {Baloch}},
  \bibinfo {author} {\bibfnamefont {P.}~\bibnamefont {Alexander}}, \bibinfo
  {author} {\bibfnamefont {Y.}~\bibnamefont {Qi}}, \bibinfo {author}
  {\bibfnamefont {J.}~\bibnamefont {Barry}}, \bibinfo {author} {\bibfnamefont
  {J.}~\bibnamefont {Melngailis}}, \bibinfo {author} {\bibfnamefont
  {L.}~\bibnamefont {Salamanca-Riba}}, \bibinfo {author} {\bibfnamefont
  {I.}~\bibnamefont {Takeuchi}}, \ and\ \bibinfo {author} {\bibfnamefont
  {J.}~\bibnamefont {Cumings}},\ }\href {\doibase
  http://dx.doi.org/10.1021/nl9036406} {\bibfield  {journal} {\bibinfo
  {journal} {Nano Letters}\ }\textbf {\bibinfo {volume} {10}},\ \bibinfo
  {pages} {1219} (\bibinfo {year} {2010})}\BibitemShut {NoStop}%
\bibitem [{\citenamefont {Summers}\ and\ \citenamefont
  {Wun-Fogle}(2007)}]{Summers:2007}%
  \BibitemOpen
  \bibfield  {author} {\bibinfo {author} {\bibfnamefont {T.~A.}\ \bibnamefont
  {Summers}, \bibfnamefont {E.~M.and~Lograsso}}\ and\ \bibinfo {author}
  {\bibfnamefont {M.}~\bibnamefont {Wun-Fogle}},\ }\href {\doibase
  10.1007/s10853-007-2096-6} {\bibfield  {journal} {\bibinfo  {journal} {J.
  Mat. Sci.}\ }\textbf {\bibinfo {volume} {42}},\ \bibinfo {pages} {9582}
  (\bibinfo {year} {2007})}\BibitemShut {NoStop}%
\bibitem [{\citenamefont {Thiele}(1973)}]{Thiele:1973}%
  \BibitemOpen
  \bibfield  {author} {\bibinfo {author} {\bibfnamefont {A.~A.}\ \bibnamefont
  {Thiele}},\ }\href {\doibase 10.1103/PhysRevLett.30.230} {\bibfield
  {journal} {\bibinfo  {journal} {Phys. Rev. Lett.}\ }\textbf {\bibinfo
  {volume} {30}},\ \bibinfo {pages} {230} (\bibinfo {year} {1973})}\BibitemShut
  {NoStop}%
\bibitem [{\citenamefont {Yu}\ \emph {et~al.}(2013)\citenamefont {Yu},
  \citenamefont {Han}, \citenamefont {Yoo}, \citenamefont {Lee}, \citenamefont
  {Choi}, \citenamefont {Jung}, \citenamefont {Lee}, \citenamefont {Im},
  \citenamefont {Fischer},\ and\ \citenamefont {Kim}}]{Yu:2006}%
  \BibitemOpen
  \bibfield  {author} {\bibinfo {author} {\bibfnamefont {Y.-S.}\ \bibnamefont
  {Yu}}, \bibinfo {author} {\bibfnamefont {D.-S.}\ \bibnamefont {Han}},
  \bibinfo {author} {\bibfnamefont {M.-W.}\ \bibnamefont {Yoo}}, \bibinfo
  {author} {\bibfnamefont {K.-S.}\ \bibnamefont {Lee}}, \bibinfo {author}
  {\bibfnamefont {Y.-S.}\ \bibnamefont {Choi}}, \bibinfo {author}
  {\bibfnamefont {H.}~\bibnamefont {Jung}}, \bibinfo {author} {\bibfnamefont
  {J.}~\bibnamefont {Lee}}, \bibinfo {author} {\bibfnamefont {M.-Y.}\
  \bibnamefont {Im}}, \bibinfo {author} {\bibfnamefont {P.}~\bibnamefont
  {Fischer}}, \ and\ \bibinfo {author} {\bibfnamefont {S.-K.}\ \bibnamefont
  {Kim}},\ }\href {\doibase http://dx.doi.org/10.1038/srep01301} {\bibfield
  {journal} {\bibinfo  {journal} {Sci. Rep.}\ }\textbf {\bibinfo {volume}
  {3}},\ \bibinfo {pages} {1301} (\bibinfo {year} {2013})}\BibitemShut
  {NoStop}%
\bibitem [{\citenamefont {Sukhostavets}\ \emph {et~al.}(2013)\citenamefont
  {Sukhostavets}, \citenamefont {Gonz\'alez},\ and\ \citenamefont
  {Guslienko}}]{Sukhostavets:2013}%
  \BibitemOpen
  \bibfield  {author} {\bibinfo {author} {\bibfnamefont {O.~V.}\ \bibnamefont
  {Sukhostavets}}, \bibinfo {author} {\bibfnamefont {J.}~\bibnamefont
  {Gonz\'alez}}, \ and\ \bibinfo {author} {\bibfnamefont {K.~Y.}\ \bibnamefont
  {Guslienko}},\ }\href {\doibase 10.1103/PhysRevB.87.094402} {\bibfield
  {journal} {\bibinfo  {journal} {Phys. Rev. B}\ }\textbf {\bibinfo {volume}
  {87}},\ \bibinfo {pages} {094402} (\bibinfo {year} {2013})}\BibitemShut
  {NoStop}%
\bibitem [{\citenamefont {Asmat-Uceda}\ \emph {et~al.}(2015)\citenamefont
  {Asmat-Uceda}, \citenamefont {Cheng}, \citenamefont {Wang}, \citenamefont
  {Clarke}, \citenamefont {Tchernyshyov},\ and\ \citenamefont
  {Buchanan}}]{Asmat:2015}%
  \BibitemOpen
  \bibfield  {author} {\bibinfo {author} {\bibfnamefont {M.}~\bibnamefont
  {Asmat-Uceda}}, \bibinfo {author} {\bibfnamefont {X.}~\bibnamefont {Cheng}},
  \bibinfo {author} {\bibfnamefont {X.}~\bibnamefont {Wang}}, \bibinfo {author}
  {\bibfnamefont {D.~J.}\ \bibnamefont {Clarke}}, \bibinfo {author}
  {\bibfnamefont {O.}~\bibnamefont {Tchernyshyov}}, \ and\ \bibinfo {author}
  {\bibfnamefont {K.~S.}\ \bibnamefont {Buchanan}},\ }\href {\doibase
  http://dx.doi.org/10.1063/1.4916610} {\bibfield  {journal} {\bibinfo
  {journal} {J. Appl. Phys.}\ }\textbf {\bibinfo {volume} {117}},\ \bibinfo
  {pages} {123916} (\bibinfo {year} {2015})}\BibitemShut {NoStop}%
\bibitem [{\citenamefont {Shibata}\ \emph {et~al.}(2006)\citenamefont
  {Shibata}, \citenamefont {Nakatani}, \citenamefont {Tatara}, \citenamefont
  {Kohno},\ and\ \citenamefont {Otani}}]{Shibata:2006}%
  \BibitemOpen
  \bibfield  {author} {\bibinfo {author} {\bibfnamefont {J.}~\bibnamefont
  {Shibata}}, \bibinfo {author} {\bibfnamefont {Y.}~\bibnamefont {Nakatani}},
  \bibinfo {author} {\bibfnamefont {G.}~\bibnamefont {Tatara}}, \bibinfo
  {author} {\bibfnamefont {H.}~\bibnamefont {Kohno}}, \ and\ \bibinfo {author}
  {\bibfnamefont {Y.}~\bibnamefont {Otani}},\ }\href {\doibase
  10.1103/PhysRevB.73.020403} {\bibfield  {journal} {\bibinfo  {journal} {Phys.
  Rev. B}\ }\textbf {\bibinfo {volume} {73}},\ \bibinfo {pages} {020403}
  (\bibinfo {year} {2006})}\BibitemShut {NoStop}%
\end{thebibliography}%

\end{document}